\documentstyle[prl,psfig,aps,floats]{revtex}  
\input psfig

\def\bb{$b\bar{b}$}
\def\cc{$c\bar{c}$}

\def\jpsi{$J/ \psi$} 
\def\psipr{$\psi'$} 

\def\pt{$p_T$}
\def\mt{$m_t$}

\def\ptmu{$p_T^{\mu}$}

\def\dr{$\Delta$$\cal R$}
\def\psimumu{$ J/ \psi \rightarrow \mu^+ \mu^- $}
\def\Buu{$ B^0 \rightarrow  \mu^+ \mu^- $}
\def\bsuu{$ b \rightarrow X_s \mu^+ \mu^- $}
\def\bduu{$ b \rightarrow X_d \mu^+ \mu^- $}

\def\mass{$M_{\mu\mu}$}

\def\ptmumu{$p_{T}^{\mu\mu}$}

\def\bb{$b\bar{b}$}

\begin{document}

\title { \large \bf Search for the Decay \bsuu\ } 

%
\author{                                                                      
B.~Abbott,$^{30}$                                                             
M.~Abolins,$^{27}$                                                            
B.S.~Acharya,$^{45}$                                                          
I.~Adam,$^{12}$                                                               
D.L.~Adams,$^{39}$                                                            
M.~Adams,$^{17}$                                                              
S.~Ahn,$^{14}$                                                                
H.~Aihara,$^{23}$                                                             
G.A.~Alves,$^{10}$                                                            
N.~Amos,$^{26}$
% ...
\\                                                                            
\vskip 0.50cm                                                                 
\centerline{(D\O\ Collaboration)}                                             
\vskip 0.50cm                                                                 
}                                                                             
\address{                                                                     
\centerline{$^{1}$Universidad de los Andes, Bogot\'{a}, Colombia}             
\centerline{$^{2}$University of Arizona, Tucson, Arizona 85721}               
\centerline{$^{3}$Boston University, Boston, Massachusetts 02215}             
\centerline{$^{4}$Brookhaven National Laboratory, Upton, New York 11973}      
\centerline{$^{5}$Brown University, Providence, Rhode Island 02912}           
\centerline{$^{6}$Universidad de Buenos Aires, Buenos Aires, Argentina}       
\centerline{$^{7}$University of California, Davis, California 95616}          
\centerline{$^{8}$University of California, Irvine, California 92697}         
\centerline{$^{9}$University of California, Riverside, California 92521}      
\centerline{$^{10}$LAFEX, Centro Brasileiro de Pesquisas F{\'\i}sicas,        
                  Rio de Janeiro, Brazil}                                     
\centerline{$^{11}$CINVESTAV, Mexico City, Mexico}                            
\centerline{$^{12}$Columbia University, New York, New York 10027}             
\centerline{$^{13}$Delhi University, Delhi, India 110007}                     
\centerline{$^{14}$Fermi National Accelerator Laboratory, Batavia,            
                   Illinois 60510}                                            
\centerline{$^{15}$Florida State University, Tallahassee, Florida 32306}      
\centerline{$^{16}$University of Hawaii, Honolulu, Hawaii 96822}              
\centerline{$^{17}$University of Illinois at Chicago, Chicago,                
                   Illinois 60607}                                            
\centerline{$^{18}$Indiana University, Bloomington, Indiana 47405}            
\centerline{$^{19}$Iowa State University, Ames, Iowa 50011}                   
\centerline{$^{20}$Korea University, Seoul, Korea}                            
\centerline{$^{21}$Kyungsung University, Pusan, Korea}                        
\centerline{$^{22}$Institute of Nuclear Physics, Krak\'ow, Poland}            
\centerline{$^{23}$Lawrence Berkeley National Laboratory and University of    
                   California, Berkeley, California 94720}                    
\centerline{$^{24}$Louisiana Tech University, Ruston, Louisiana 71272}        
\centerline{$^{25}$University of Maryland, College Park, Maryland 20742}      
\centerline{$^{26}$University of Michigan, Ann Arbor, Michigan 48109}         
\centerline{$^{27}$Michigan State University, East Lansing, Michigan 48824}   
\centerline{$^{28}$Moscow State University, Moscow, Russia}                   
\centerline{$^{29}$University of Nebraska, Lincoln, Nebraska 68588}           
\centerline{$^{30}$New York University, New York, New York 10003}             
\centerline{$^{31}$Northeastern University, Boston, Massachusetts 02115}      
\centerline{$^{32}$Northern Illinois University, DeKalb, Illinois 60115}      
\centerline{$^{33}$Northwestern University, Evanston, Illinois 60208}         
\centerline{$^{34}$University of Notre Dame, Notre Dame, Indiana 46556}       
\centerline{$^{35}$University of Oklahoma, Norman, Oklahoma 73019}            
\centerline{$^{36}$University of Panjab, Chandigarh 16-00-14, India}          
\centerline{$^{37}$Institute for High Energy Physics, 142-284 Protvino,       
                   Russia}                                                    
\centerline{$^{38}$Purdue University, West Lafayette, Indiana 47907}          
\centerline{$^{39}$Rice University, Houston, Texas 77005}                     
\centerline{$^{40}$Universidade do Estado do Rio de Janeiro, Brazil}          
\centerline{$^{41}$University of Rochester, Rochester, New York 14627}        
\centerline{$^{42}$CEA, DAPNIA/Service de Physique des Particules,            
                   CE-SACLAY, Gif-sur-Yvette, France}                         
\centerline{$^{43}$Seoul National University, Seoul, Korea}                   
\centerline{$^{44}$State University of New York, Stony Brook,                 
                   New York 11794}                                            
\centerline{$^{45}$Tata Institute of Fundamental Research,                    
                   Colaba, Mumbai 400005, India}                              
\centerline{$^{46}$University of Texas, Arlington, Texas 76019}               
\centerline{$^{47}$Texas A\&M University, College Station, Texas 77843}       
}                                                                             

\maketitle

\begin{abstract}

We have searched for the flavor-changing neutral current decay
\bsuu\ in $p\bar p$  collisions at 
$\sqrt{s} =$1.8 TeV with the D\O\ detector at Fermilab. 
We determine the 90\% confidence level limit for the branching fraction 
to be $B ( b \rightarrow X_s \mu^+ \mu^- $) $<$ 3.2$\times$10$^{-4}$. 
We argue that this limit is more stringent than the best published limit
on this decay rate.

\end{abstract}

\twocolumn
\nopagebreak

    In the Standard Model of electroweak interactions (SM) the decay 
processes \bsuu\ (where $X_s$ stands for a hadronic  state containing
a strange  quark) and \Buu\ are forbidden at tree level
and are possible only through loop diagrams. The largest contributions
to the branching fraction for these processes
come from diagrams involving the top quark and therefore 
the predicted branching fractions depend on the top quark mass.
The ``rare" flavor-changing neutral current (FCNC)
decays are expected to be more frequent for $b$ quarks
than for strange quarks. 
The $b \rightarrow c$
transition is suppressed by $|V_{bc}|$ = ${\cal O} (10^{-1})$ while loop 
corrections 
are large due to $|V_{tb}|$ $\approx$ 1 and the large top quark mass, \mt.
At \mt = 170 GeV/c$^2$, the 
SM expected branching fraction\cite{bsuutheor,bsuumass_ab} 
for the semi-inclusive decay \bsuu\ 
  is 6$\times10^{-6}$. 

Extensions to the minimal Standard Model which allow new particles 
contributing to the higher order corrections, 
such as fourth generation quarks, charged Higgs bosons 
or supersymmetric 
particles, provide additional possible sources of FCNC. Precision measurements
of  rare $b$ decay rates  thus extend the new physics discovery potential 
beyond direct searches. 
We describe a search for such decays in the data collected with the 
D\O\ detector\cite{dzero} during the 1994--1995 Fermilab Tevatron run.
The data correspond to a total integrated luminosity of 
$\mathcal {L}$ = 50.0 $\pm$ 2.7 pb$^{-1}$.

Dimuon events are selected by requiring 
two muons in the central muon system, at both hardware and software 
trigger levels\cite{level1}.
We select events containing an oppositely charged muon pair
with the invariant mass \mass\ $<$ 7 GeV/c$^2$, transverse momentum
\ptmumu\ $>$ 5 GeV/c, and pseudorapidity $|\eta_{\mu\mu}|$ $<$  0.6.
The muons are required to have a transverse momentum 
\ptmu\ $>$ 3.5 GeV/c and pseudorapidity 
$|\eta_\mu|$ $<$  1.0.

Both muon trajectories  are required to be consistent 
with the reconstructed vertex position and to have 
a matching track in the central detector. 
There must be appropriate energy deposition along the muons' path
in the calorimeter.
The total number of events satisfying the above criteria is 1564.

\begin{figure}[tb]
\centerline{\psfig{figure=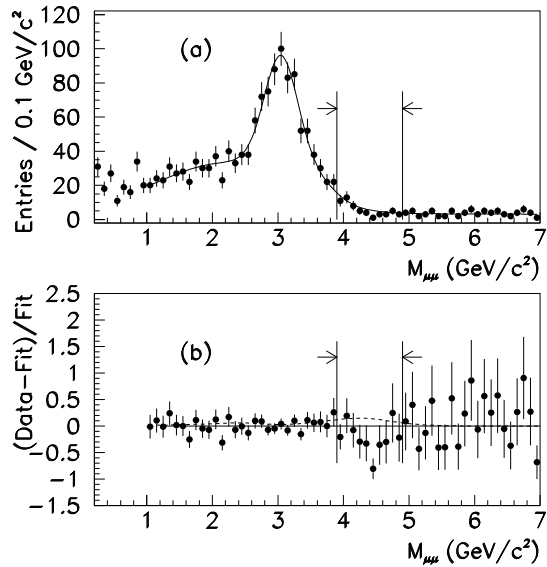,height=8cm}}
\caption{
(a) Dimuon invariant mass spectrum.
The solid line is the maximum likelihood fit of the known
physics processes to the data (see text);
(b) the data points after subtraction of the fitted values,
divided by the fitted values.
Only statistical uncertainties on the data points are shown.
The dashed line corresponds
to the 90\% confidence level upper limit for the
decay \bsuu\ obtained from the  fit.
The arrows indicate the search window for the decay \bsuu.
}
\label{rare_suu_dif_5e}
\end{figure}

The dimuon mass spectrum of these events is shown in 
Fig.~\ref{rare_suu_dif_5e}.
In addition to the \jpsi\ resonance, the major known 
sources\cite{psid0} of
dimuons for \mass\ $<$ 7 GeV/c$^2$ are:
(1) \bb\  and \cc\ events  with both heavy
quarks decaying semileptonically or with a
sequential semileptonic decay
$b \rightarrow c + \mu$, $c \rightarrow s + \mu$;
(2) the case where one muon comes from a $b$ or $c$ decay
and the other from the decay of a $\pi$ or $K$ meson, and
(3) virtual photon decays (the Drell-Yan process).

The curve in  Fig.~\ref{rare_suu_dif_5e}(a) shows the results of a maximum 
likelihood
fit of a sum of the \jpsi\ signal and processes (1)--(3) to
the dimuon mass spectrum in the range 1 $<$ \mass\ $<$ $7$ GeV/c$^2$.
By \jpsi\ signal we mean the \jpsi\ plus an admixture of \psipr.
We use the  ratio $N$(\psipr)/$N$(\jpsi) = (6$\pm$2)\%,
based on the recent results from the CDF collaboration\cite{cdfpsi,cdfpsipr},
corrected for the mass dependence of the 
kinematic acceptance for dimuons at D\O. 
The \psipr\ mass is assumed to be higher than the \jpsi\ mass
by 0.59 GeV/c$^2$\cite{pdg},
and its width is assumed to be 20\% larger than the \jpsi\ width.
The normalized \mass\ distributions for processes (1)--(3) 
were obtained
by fitting the corresponding Monte Carlo (MC) simulated spectra.
We use a dimuon mass resolution function parametrized by a 
superposition of two Gaussians, with the shape determined  by 
fitting  MC simulated events of the decay  \psimumu\
and the mass scale allowed to vary.
For a five parameter fit to 60 bins in Fig.~\ref{rare_suu_dif_5e}(a), 
we obtain $\chi^2$(55)=56. The overall quality of the fit 
gives us confidence in the simulation of the dimuon mass 
resolution and the contributing physics processes.

The search for the decay \bsuu\ is performed in the 
mass window 3.9 $<$ \mass\ $<$ $4.9$ GeV/c$^2$. 
This mass range is above most of the known sources of dimuons and hence
constitutes the region of maximum sensitivity to the
\bsuu\ decay.
Although we do not identify the strange particle among the hadrons
originating from the $b$ decay, we assume that the decay \bsuu\
dominates over the corresponding CKM-suppressed decay \bduu.  

We observe 56 events in our search window, where 68$\pm$2(stat.)$\pm$4(syst.) 
are expected from the fit.
We thus find no evidence for an excess of events to be attributed to
the decay \bsuu.
To estimate the systematic error, we have performed alternative fits, changing 
within their uncertainties the width and skewness of the dimuon mass 
resolution function and mass scale, as well as the mix of backgrounds.
MC studies of 10,000 simulated experiments,
with the background composition as
obtained in the fit to the data,
indicate that the probability of obtaining 56 events 
in a given experiment where 68$\pm$5 events are expected  is 12\%. 
The contribution to the $\chi^2$ from the 10 bins within the search window is
7.3.
                                                   
We follow two independent ways of relating the observed number of events
in the search window to the expected $b$ quark yield.
First, we use the absolute normalization to the inclusive 
$b$ quark production cross section. The result depends on
the assumed $b$ cross section, on the $b$ production model and
on the estimates for the trigger and offline reconstruction efficiencies.
A similar method has been employed previously by the UA1 Collaboration\cite{UA1}.
In the alternative approach, we normalize to the observed \jpsi\ signal,
where the uncertainties in the variables used in the denominator
of Equation (1) largely cancel. The result of the latter method
depends on the knowledge of the fraction of \jpsi\ events that originate
from $b$ decays, and on the branching fraction 
for the decay sequence
$b \rightarrow X_s  J/ \psi,J/ \psi \rightarrow \mu \mu$.

In the first approach, the branching fraction for the decay 
\bsuu\ is given by
\begin{equation}
B (b \rightarrow X_s \mu^+ \mu^-) = \frac {N}
{2 \cdot \sigma (b) \cdot \mathcal {L} \cdot
\epsilon }\ ,
\end{equation}
where $N$ is the number of events due to this decay,
$\sigma (b)$ denotes the inclusive $b$ quark production cross section for
$p_T(b)$ $>$ 6 GeV/c and rapidity  $|y(b)|<1$, 
and $\epsilon$ is the combined kinematic
acceptance and trigger and offline reconstruction
efficiency.
                                                
\begin{figure}[tb]
\centerline{\psfig{figure=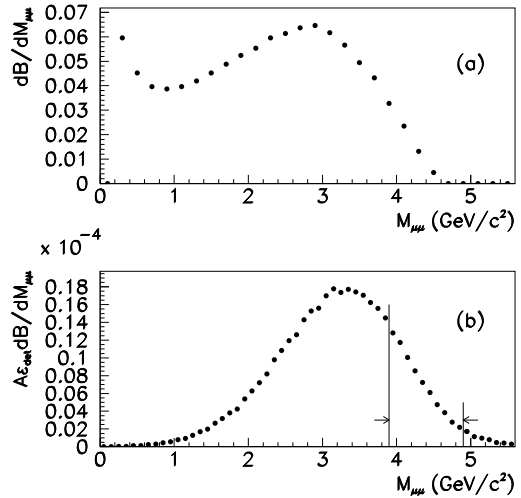,height=7.cm}}
\caption{ 
(a) The calculated differential branching fraction for the decay \bsuu,
from Ref. [2] as a function of \mass\ (multiplied by 0.2 GeV/c$^2$); 
(b) the same differential branching fraction modified by the response of 
the D\O\ detector (multiplied by 0.1 GeV/c$^2$). 
The arrows indicate the search window used in this analysis.}
\label{bxuueff}
\end{figure}

The inclusive $b$ quark production cross section for 
$p_T(b)$ $>$ 6 GeV/c and $|y(b)|<1$
is $\sigma (b)$ = 7.2  $\pm$ 1.8  $\mu$b.  
This estimate results from a fit to a compilation of D\O\ 
measurements  \cite{bcross}
of the integrated, inclusive $b$ production cross section
at $p_T(b)$ $>$ $p_T^{\text {min}}$.

The calculation of $\epsilon$ proceeds by multiplying the theoretical 
mass spectrum\cite{bsuumass_ab}, normalized to unity,
by the mass-dependent detection efficiency,  convoluting
the detector resolution, and integrating
the resulting distribution over the search window.
The simulated mass spectrum including the detector response
is compared to the input distribution from Ref.\cite{bsuumass_ab}
in Fig.~\ref{bxuueff}.

The mass-dependent dimuon detection efficiency is determined from 
events with $b$ quarks generated with the  
{\footnotesize ISAJET} program\cite{isajet} in the lowest  order 
QCD approximation.
Quarks that satisfy the above kinematic requirements
are fragmented according to the Peterson fragmentation
model \cite{Peterson}.
We adjust the value of the fragmentation parameter $\epsilon_b$ to obtain 
the dimuon
transverse momentum spectrum that matches the \pt\ spectrum of \jpsi\
coming from $b$ quark decay, measured by the CDF Collaboration\cite{cdfpsi}.

To expedite the simulation procedure, preselection cuts 
(called  $\mathcal K$) of   \ptmu\ $>$ 3 GeV/c and $|\eta_\mu|$ $<$  1.0 
are applied to both muons. The acceptance $A(M_{\mu \mu })$ 
for this preselection increases with \mass. 
It is determined by studying MC samples of the
decay \bsuu\ generated at various values of \mass.
At the parton level, the decay \bsuu\ has the same final state
as the $b$ decay to \jpsi, as illustrated in the diagrams below:

\centerline{\psfig{figure=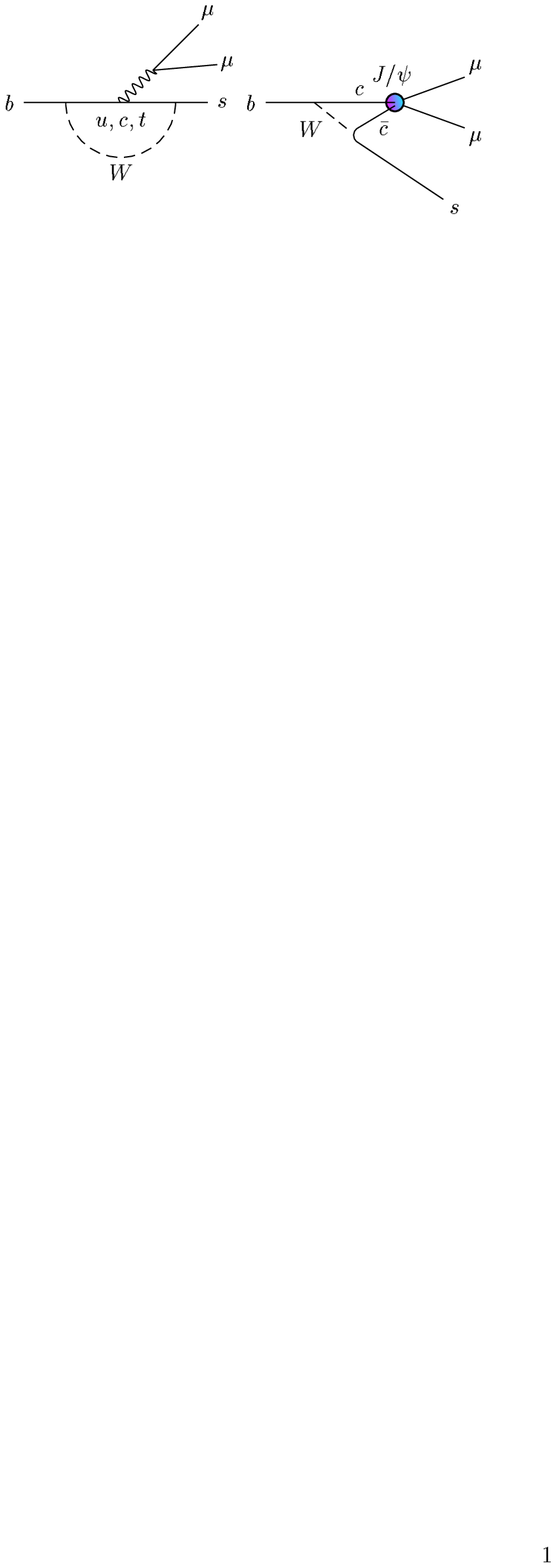,height=6.cm}}

Therefore, we use ``$J/\psi+ X_s$", simulated according to Ref.\cite{isajet},
as a model for the final state of the decay \bsuu\
by substituting various discrete values for the \jpsi\ mass.
At \mass\ = 4.1 GeV/c$^2$ we remove multibody channels, keeping 
the channels  $J/\psi K$ and  $J/\psi K^*$  with the relative rate 1:2.
We find the resulting acceptance insensitive to the number of
channels included and their relative rates. 
We also calculate the acceptance for the exclusive $B$ meson 
decay to a muon pair.
The results
are fitted with the form $A$(\mass ) = 2.9$\times 10^{-3}\times$\mass $^{2.64}$.
The uncertainty on  $A(4.1)$, taken from the difference between
its values at $\epsilon_b$=0.02 and at $\epsilon_b$=0.006, is 20\%.
The uncertainty of the ratio of  $A(4.1)$ to  $A(3.1)$  is 5\%.

Events that satisfy the muon pseudorapidity and momentum cuts 
($\mathcal K$) are passed through a detector simulation, trigger 
simulation, and offline reconstruction programs. 
We find the effect of the trigger and reconstruction efficiency 
for those events to be independent of \mass.
The trigger efficiency, corrected for effects not included in
the simulation, is 0.052 $\pm$ 0.005. The efficiency of the
offline selection cuts, 0.19 $\pm$ 0.03, has been obtained by comparing 
the total number of \jpsi\ events passing the dimuon trigger, 3310 $\pm$ 500,
to the number of triggered \jpsi\ events that satisfy our  offline 
selection cuts, 633 $\pm$ 45. The product of the trigger and offline selection
efficiencies is $\epsilon_{\text {det}}$ =(1.0 $\pm$ 0.2)$\times$10$^{-2}$.

The integral of the spectrum in Fig.~\ref{bxuueff}(b)
over the search window is
$\epsilon$ = (7.0 $\pm$ 2.0)$\times$10$^{-5}$. 
The acceptance of the kinematic selection alone, including the selection 
($\mathcal K$) and the mass cut, is (7.0 $\pm$ 1.4)$\times$10$^{-3}$.

The \jpsi\ signal allows an alternative normalization,
to the \jpsi\ yield due to the $b$ quark decay, $N_{b \rightarrow  J/\psi}$.
The ratio of the kinematic acceptance $A$(\mass ) integrated over 
the search window to $A(3.1)$ is $\alpha$ = 0.123 $\pm$ 0.006. 
From the total number of \jpsi\ events in our dimuon sample,  633 $\pm$ 45, 
and the fraction  of \jpsi\ events originating from $b$ decay, 
$f_b$ = 0.31 $\pm$ 0.03\cite{cdfpsi,fb}, the number of \jpsi\ events
coming from $b$ decays is $N_{b \rightarrow  J/\psi}$=196 $\pm$ 23. 
With the branching fraction for the decay sequence
$b \rightarrow X_s  J/\psi, J/\psi \rightarrow \mu \mu$,  
$B(b \rightarrow J/\psi \rightarrow \mu \mu)$ = 
(7.0$\pm$0.6) $\times$ 10$^{-4}$\cite{pdg}, we have:

\begin{equation}
B (b \rightarrow X_s \mu^+ \mu^-) = \frac {N}
{N_{b \rightarrow  J/\psi}  / B(b \rightarrow J/\psi \rightarrow \mu \mu)   \cdot \alpha   }.
\end{equation}

The respective results for the denominators in Eqs. (1) and (2), 
(5.0$ \pm$ 1.9)$\times$10$^{4}$ and (3.4 $\pm$ 0.5)$\times$10$^{4}$, 
are consistent.
Using the latter by virtue of its smaller systematic error,
we obtain a 90\% confidence level limit of
$B (b  \rightarrow  X_s \mu^{+}\mu^{-} ) < 3.2 \times 10 ^{-4}$.
To derive the limit, we apply a Bayesian approach in which the
observed number of events is compared to the number of background
events in the region of interest. We assume Poisson statistics
for the signal and background and account for uncertainty in the 
background and in the estimates of the total cross section and of the dimuon
detection efficiency.
We have found the results to be stable with respect to the choice
of the search window by varying the lower and upper limits of the 
window within $\pm$ 50 MeV/c$^2$ and $\pm$ 200 MeV/c$^2$,
repsectively.

 The best published limit for this decay ( 5$\times$10$^{-5}$ )
was set by the UA1 Collaboration\cite{UA1}.
We have attempted to reproduce the UA1 limit 
and to make the cross-check between their
quoted efficiency and their \jpsi\ signal\cite{UA1eff}. We have 
failed to reconcile the two. Instead of 
the quoted efficiency of  0.011 we obtain\cite{UA1eff_d0} 
$\epsilon_{\rm UA1}$  $\approx$ 5.8$\times$10$^{-4}$ --
lower by a factor $\approx$ 20. 
Differences between the theoretical dimuon
mass distributions for the decay \bsuu,  or in the
versions of the {\footnotesize ISAJET} program that were used here 
and in Ref.\cite{UA1}, cannot account for such a large disparity 
in the results.
Using our estimates of their efficiency we obtain 
$\approx$1$\times$10$^{-3}$ as 
the upper limit on
$B (b \rightarrow X_s \mu^+ \mu^- $) from their experiment.

For the exclusive decay \Buu\
(an unseparated mixture of $B_d$ and $B_s$ decays)
we define the search window
as 4.8 $<$ \mass\ $<$ $5.8$ GeV/c$^2$, resulting in the
maximum sensitivity to the signal. The acceptance for this
mass window is 0.60 $\pm$ 0.03.
In this process, the two muons are
expected to carry a large fraction of the energy in a cone 
around the direction
of the parent $b$ quark. To reduce background, we select events whose
energy deposition in the calorimeter in a cone 
around each muon of radius \dr $=$ 0.4 
in the pseudorapidity -- azimuthal angle space 
is less than 8 GeV. The acceptance of the isolation requirement is
0.80 $\pm$ 0.03.

\begin{figure}[tb]
\centerline{\psfig{figure=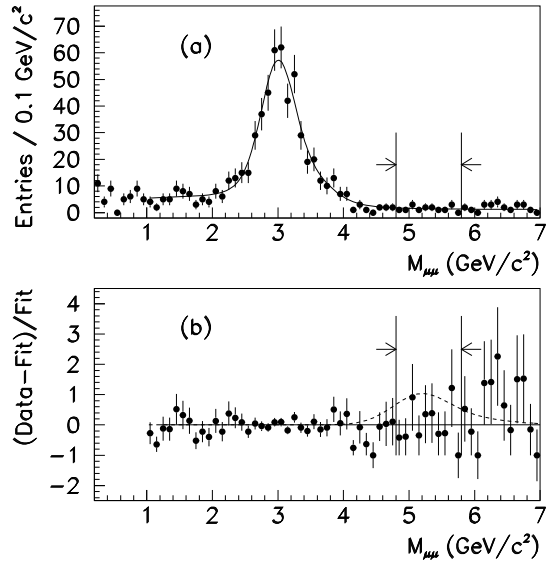,height=8.cm}}
\caption{
(a) The invariant mass distribution for isolated dimuons.
The solid line is the maximum likelihood fit of the known
physics processes to the data (see text);
(b) the data points after fit subtraction  divided by the fitted values.
The dashed line corresponds
to the 90\% confidence level upper limit for the
decay \Buu\ obtained from the  fit.
The arrows indicate the search window for the decay \Buu.}
\label{rare_uu_dif_5e}
\end{figure}

The mass spectrum for isolated dimuons is shown in
Fig.~ \ref{rare_uu_dif_5e}(a).
We find 15 events in the search region.
From  a fit to the sum of the \jpsi\ signal and processes
(1)--(3), the background
in the search window is estimated to be 15 $\pm$ 2 events.

\nopagebreak

The $B^0$  production cross section at 
$p_T(B)$ $>$ 6 GeV and $|y(B)|<1$, measured by the CDF collaboration 
\cite{CDF}, is 2.39 $\pm$ 0.54 $\mu$b.
The product of the acceptance
for kinematic and geometric restrictions on the two muons
coming from the  decay of a $B^0$ meson, the trigger and
offline reconstruction efficiency, and the mass and isolation cuts
is $\epsilon$ = (1.4 $\pm$ 0.35)$\times$10$^{-3}$.
We obtain  a 90\% ~confidence level limit of 
$B(B^0  \rightarrow \mu^{+}\mu^{-} ) < 4.0 \times 10 ^{-5}$.
The  best published limit for this decay 
is 1.6$\times$10$^{-6}$, set by the CDF Collaboration in Ref.\cite{CDF}. 
The SM prediction   is 1.5$\times$10$^{-10}$.
We estimate\cite{UA1eff_uu} that the limit published by the UA1 collaboration 
in Ref.\cite{UA1} should be shifted upward by about a factor of four. 

    In conclusion, we have conducted a search for the
FCNC decays \bsuu\ and \Buu.
We find no evidence for either decay. For the semi-inclusive decay 
\bsuu\ we set a 90\% confidence level limit of
$B (b  \rightarrow X_s \mu^{+}\mu^{-} ) < 3.2 \times 10 ^{-4}$.
In view of our observations we conclude\cite{CLEO} that this limit 
is more stringent than the best published limit on this decay rate.
The SM prediction   is 6$\times$10$^{-6}$.

%
We thank the staffs at Fermilab and collaborating institutions for their
contributions to this work, and acknowledge support from the 
Department of Energy and National Science Foundation (U.S.A.),  
Commissariat  \` a L'Energie Atomique (France), 
State Committee for Science and Technology and Ministry for Atomic 
   Energy (Russia),
CNPq (Brazil),
Departments of Atomic Energy and Science and Education (India),
Colciencias (Colombia),
CONACyT (Mexico),
Ministry of Education and KOSEF (Korea),
CONICET and UBACyT (Argentina),
and CAPES (Brazil).
%



\begin{thebibliography}{10}
\bibitem{cdfpsipr}
CDF has measured the ratio of cross section times branching fraction
for \jpsi\ and \psipr\ as a function of \pt. Taking the CDF measurement
at our mean dimuon \pt\ and 
 accounting for the \mass\  dependence of the dimuon 
detection efficiency at D\O, we
obtain the expected ratio of the number of events of \psipr\ and 
\jpsi\ in our sample to be $\approx$ 0.06$\pm$0.02.
\bibitem{fb}
The value of $f_b$ used in this analysis is an interpolation of CDF 
measurements at  \ptmumu\ = 12 GeV/c.
\bibitem{UA1eff}

With $\sigma (b)$ = 2.5 $\mu$b and $\mathcal {L}$ = 5.3  pb$^{-1}$,
the number of $J/\psi$ events from $b$ decays in the dimuon channel 
at $\sqrt{s}$ = 630 GeV is
$\approx$19,000. UA1 observes $\approx$250 $J/\psi$ events,
$\approx$78 of them originating from $b$ decays.
Thus, the apparent efficiency is $\approx$0.0041. This is inconsistent
with the value 0.073 (=14713/200000) quoted by UA1 for the efficiency
averaged over a spectrum that peaks below 3 GeV/c$^2$.
\bibitem{UA1eff_uu}
From our {\footnotesize ISAJET} simulation 
we obtain the UA1 kinematic acceptance for the fragmentation process
$b \rightarrow B^o$,   followed by the decay 
\Buu, to be 0.14.
With their mass independent detection efficiency of $\approx$0.1, the overall
effect of the kinematic acceptance and detection efficiency for this decay 
then is 0.014. UA1 quotes an efficiency of 11465/200000 = 0.057.
\end{thebibliography}
\end{document}